# ACE-Net: AutofoCus-Enhanced Convolutional Network for Field Imperfection Estimation with application to high b-value spiral Diffusion MRI

Mengze Gao1, Zachary Shah2, Xiaozhi Cao2, Nan Wang2, Daniel Abraham2, Kawin Setsompop1,2

1 Department of Radiology, Stanford University, Stanford, CA, USA
2 Department of Electrical Engineering, Stanford University, Stanford, CA, USA

Introduction:

Spatiotemporal field variations from $B_0$-inhomogeneity and eddy-currents are detrimental to rapid image-encoding schemes such as spiral, EPI and 3D-cones. Incorporation of field-imperfections into the reconstruction can mitigate the undesirable image artifacts. However, high-quality estimates of these imperfections require lengthy calibration scans and/or external field-probe(1), with added confound that they can also change due to scanner heating and subject physiological and involuntary motions. To address this, data-driven field-imperfection estimation methods have been developed to estimate $B_0$-inhomogeneity(2-4), and spatio-temporal diffusion eddy current in EPI(5). Building on these works, we developed 'ACE-Net' (Autofocus-Enhanced Convolutional Network for Field Imperfection Estimation) to accurately estimate spatiotemporal $B_0$ and eddy-fields. Specifically, *ACE-Net* combines autofocus-metrics (2,3) with physic-informed deep-learning(4), while leveraging a compact basis representation of field imperfections (5). An unrolled-network build upon ACE-Net was also developed to robustify the approach against hallucination. We demonstrate our approach in high b-value spiral diffusion application.

Methods:

The description of ACE-Net is provided below in the context of diffusion spiral imaging, where the static $B_0$-inhomogeneity is first estimated from a non-diffusion-weighted image (b=0), and the remaining spatiotemporal field variations from eddy-current and dynamic $B_0$ changes are estimated in a subsequent step *per* diffusion direction.

**Static $B_0$ inhomogeneity**:

Autofocusing works by minimizing a blurring metric ($I_{auto}$) through a range of off-resonance frequencies:

$$I_{auto}(r,f) = \left|Im(I(r,f) \cdot exp^{-1j \cdot P(r)})\right|$$

where $I(r,f)$ is the reconstructed image at frequency-offset $f$, $P(r)$ is the background phase estimate, and $Im()$ is the imaginary operator. The performance of autofocus depends on image-SNR and off-resonance range, where a large range will result in local minimas. To improve performance, we integrated Autofocus with a tailored CNN-network shown in fig1, where the inputs are the blurred b=0 image and autofocus metrics across +-120 Hz at 10Hz increment. The autofocus metrics are generated using a truncated 12.5ms spiral data, to enable a valid search range of 40Hz. Using a CNN-based initial $B_0$-estimate ($B_0$-rough) (6), we limit the search range of $B_0$ ACE-Net to be within 40Hz of this estimate on a voxel-wise basis.

**Spatiotemporal eddy and $dB_0$ fields:**
This is modeled compactly via:

$$\phi(r,t) = \sum_k \phi_k(r)\delta_k(t)$$

Where $\phi_k(r)$ indicates spatial spherical-harmonic bases, and $\delta_k(t)$ indicates temporal bases. For the target diffusion imaging application, based on acquired field-probe data, we empirically determined that 3$^{rd}$ order spatial spherical-harmonic along with 3$^{rd}$ order temporal polynomial provide high-quality compact representation.

Given that eddy and dB$_0$ field variations are relatively small, a rough field estimation was not used in this network to supplement autofocus (Fig2a). The network's inputs are blurry DWI reconstructed with B$_0$-inhomogeneity correction based on B$_0$-estimate from the first network, slice location, and autofocus metric at center frequency. The output is the spatiotemporal field parameters (spatiotemporal coefficients). We also further augmented the proposed CNN with data-consistency term in an unroll fashion to reduce the risk of network hallucination (Fig2b).

**Data:**
All data were acquired using GE UHP 3T, with G$_{max}$/S$_{max}$ = 100/200 and 32ch head-coil. Test data on two volunteers were acquired using spiral diffusion at b=3000mm/s$^2$, res=1.5x1.5x5mm$^3$, R$_{inplane}$=3, acquired along with multi-echo GRE B$_0$-map and Skope eddy-current measurement to serve as ground truth. For validation, SENSE-based time-segmented reconstruction was performed with field-imperfections incorporated.

To train the ACE-Net networks, diverse training data that do not overlap with the testing cases were used. To train "B$_0$" ACE-Net, B$_0$ maps and spiral data were generated from previously-acquired 3D-MRF and B$_0$ map datasets across 10 volunteers. b=0 spiral data were created by incorporating into the forward model, the B$_0$ map and synthesized T$_2$-w images generated from PD and T$_2$ map.

To train the "spatiotemporal" ACE-Net, Skope measurement for diffusion-encoding at b1000 and 2000 with different mixing-times and TEs to the test b3000 case were used. Measurements from the 3 principal (Gx, Gy, Gz) encoding-directions were employed along with augmentations by adding up to 30% variations on the spatiotemporal basis coefficients. Furthermore, 32 non-principal diffusion-encoding directions were synthesized using linear eddy-current model with 30% random variations. Synthesized cases along with a clean DWI dataset from previously acquired BUDA-EPI (7) on 5 subjects were used to create corrupted spiral diffusion data. The spatiotemporal ACE-Net were trained on 1000 cases and unrolled ACE-Net we trained on 200.

Results
Fig.3 shows estimation of the B$_0$ map.
Fig.4 shows eddy-current field correction via CNN.
Figure 5 shows improved eddy-current field correction via unrolled network.

Fig.3 shows the B$_0$ estimation and correction.
Fig.4 shows eddy-current field estimation and correction via CNN and unrolled network.
Fig.5 shows DTI without/with field correction.

Discussion

Assisted with autofocus metrics, the network provides robust and promising result in both $B_0$ and eddy current field estimation.

Conclusion

ACE-Net was demonstrated to accurately estimate spatiotemporal field-imperfections in high b-value spiral diffusion. Future work will include rigorous validation across protocols and scanner platforms, with extensions to applications in EPI and motion-robust autofocus reconstruction.

Reference

1. Wilm, Bertram J., et al. "Single-shot spiral imaging enabled by an expanded encoding model: D emonstration in diffusion MRI." *Magnetic resonance in medicine* 77.1 (2017): 83-91.
2. Noll, Douglas C., et al. "Deblurring for non-2D Fourier transform magnetic resonance imaging." *Magnetic Resonance in Medicine* 25.2 (1992): 319-333.
3. Anderson III, Ashley G., Dinghui Wang, and James G. Pipe. "Off-resonance map refinement using autofocusing for spiral water-fat imaging." *Proceedings of the 24th Annual Meeting of ISMRM, Singapore*. 2016.
4. De Goyeneche Macaya, Alfredo, et al. "ResoNet: Noise-Trained Physics-Informed MRI Off-Resonance Correction." *Advances in Neural Information Processing Systems* 36 (2024).
5. Valsamis, Jake J., Paul I. Dubovan, and Corey A. Baron. "Characterization and correction of time-varying eddy currents for diffusion MRI." *Magnetic Resonance in Medicine* 87.5 (2022): 2209-2223.
6. Gao, Mengze, et al. "Sequence adaptive field-imperfection estimation (SAFE): retrospective estimation and correction of $B_1^+$ and $B_0$ inhomogeneities for enhanced MRF quantification." *arXiv preprint arXiv:2312.09488* (2023).
7. Liao, Congyu, et al. "Distortion-free, high-isotropic-resolution diffusion MRI with gSlider BUDA-EPI and multicoil dynamic B0 shimming." *Magnetic resonance in medicine* 86.2 (2021): 791-803.

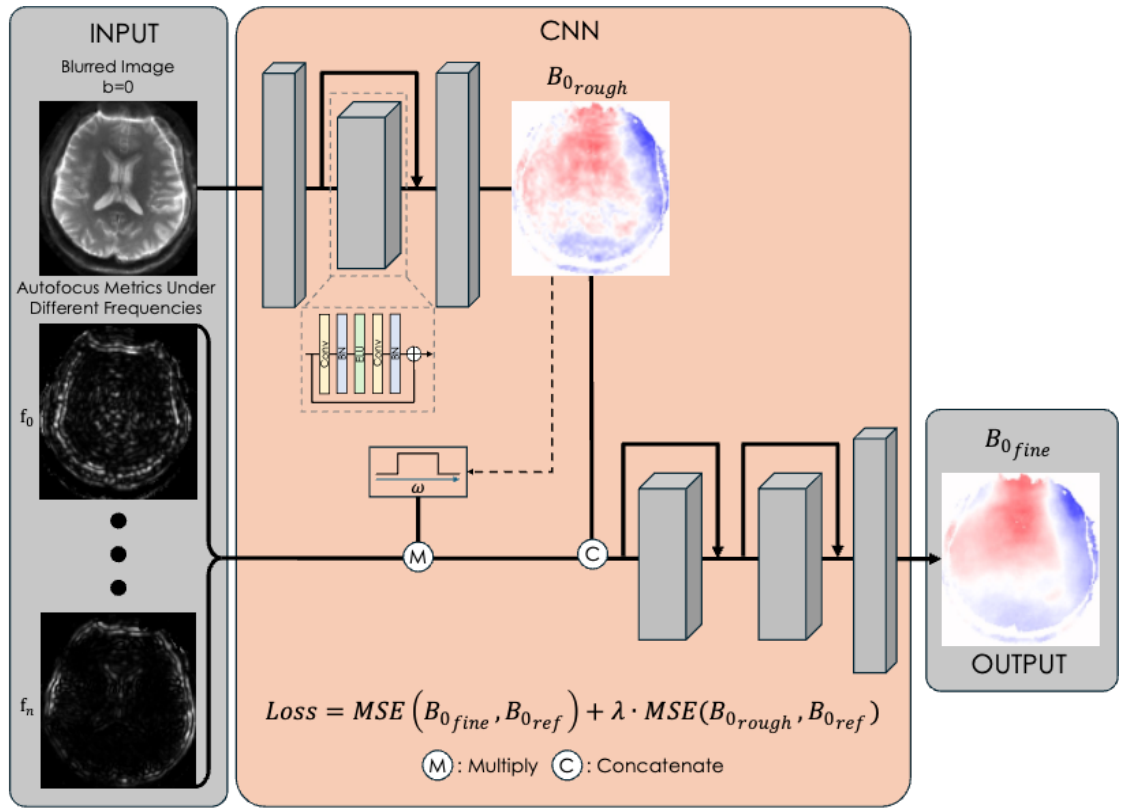

Figure 1. Network architecture for $B_0$ field estimation. The inputs to the network are diffusion-weighted images (DWI) corrupted by $B_0$ and eddy current fields, slice location, and autofocus metrics at multiple frequency offsets. The output is the estimated $B_0$ map. A frequency selection mask based on an initial rough $B_0$ estimation is used to select autofocus metrics within the appropriate frequency range. The network comprises several ResNet blocks, and the loss is computed between the ground truth and both the rough and refined B0 estimates.

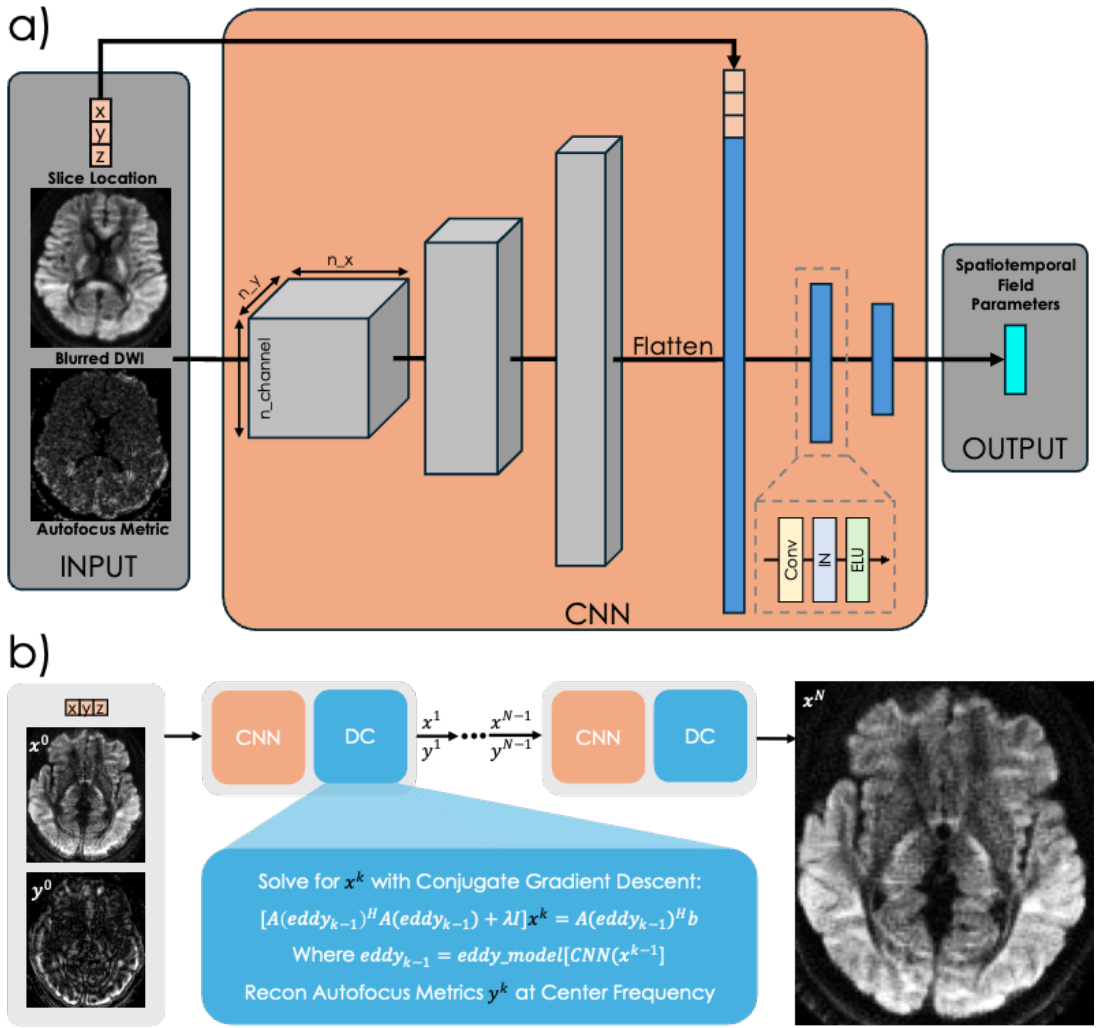

Figure 2. Network architecture for spatiotemporal eddy current and $dB_0$ field estimation. (a) The Spatiotemporal ACE-Net structure: the inputs are $B_0$-corrected DWIs, slice location, and autofocus metrics at the center frequency. The outputs are the spatiotemporal field parameters, which are further used to characterize eddy current and $dB_0$ fields. (b) The unrolled Spatiotemporal ACE-Net structure: the blurred DWIs and autofocus metrics are updated via data consistency (DC) blocks.

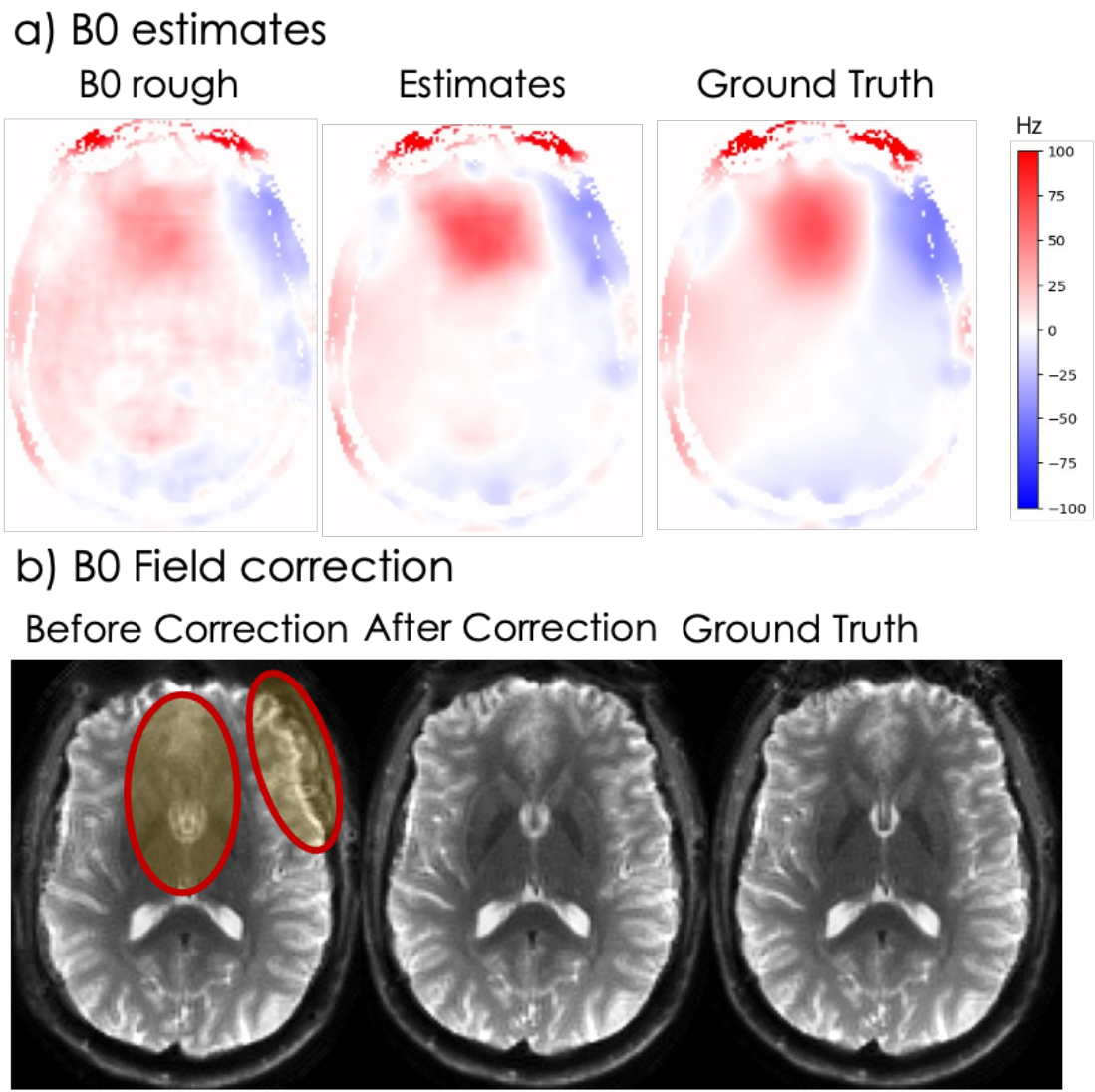

Figure 3. (a) $B_0$ field estimation results. The rough $B_0$ estimation shows high similarity to the ground truth. By utilizing autofocus metrics at different frequencies, the fine $B_0$ estimation further improves accuracy, ultimately achieving results with high correspondence to the ground truth. The ground truth $B_0$ map was acquired via multi-echo EPI. (b) After $B_0$ field correction, $B_0$ blur artifacts in certain areas are mitigated.

## a) Spatial-temporal Field Estimates

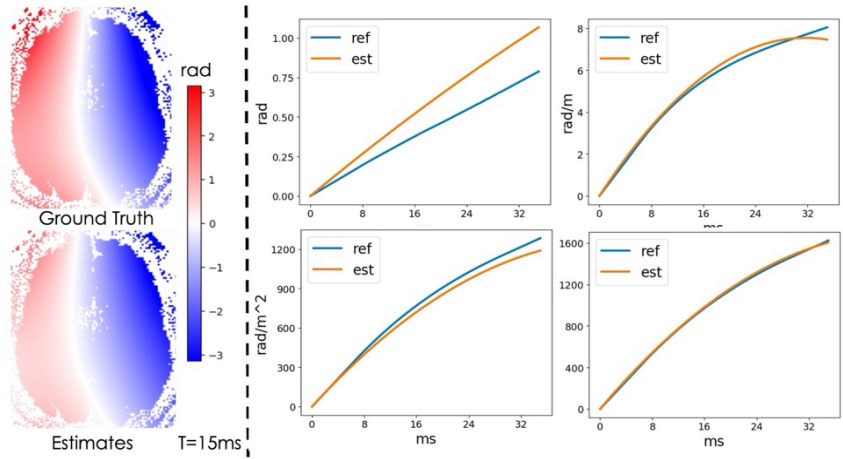

## b) Reconstruction at b=3000 s/mm² (CNN)

Before Correction   After Correction   Ground Truth

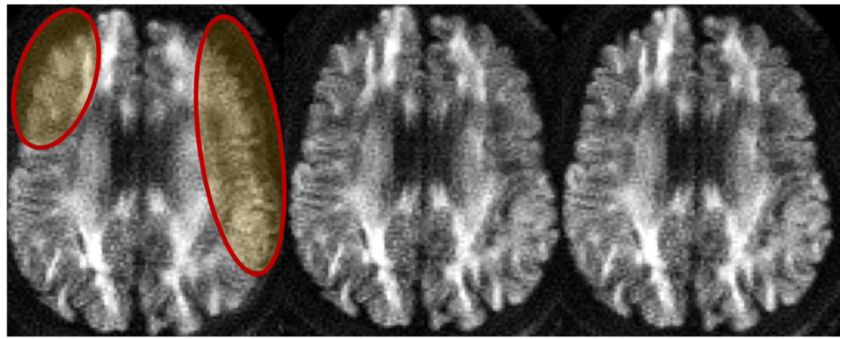

## c) Spatial-temporal Field Estimates

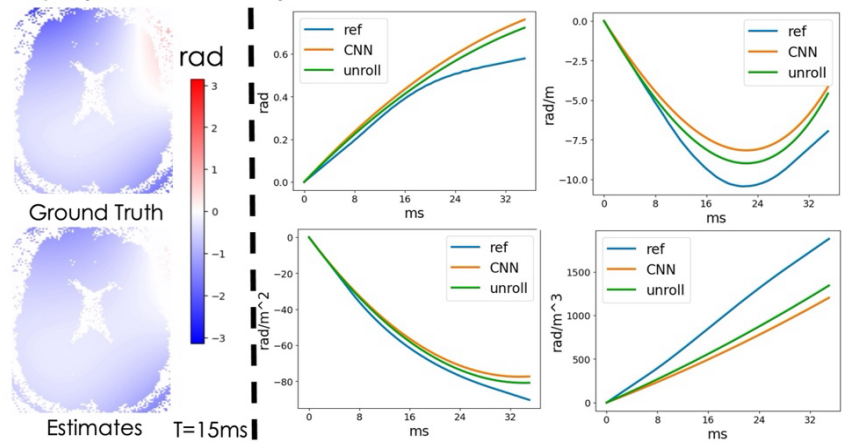

## d) Reconstruction at b=3000 s/mm² (Unroll)

Before Correction   After Correction

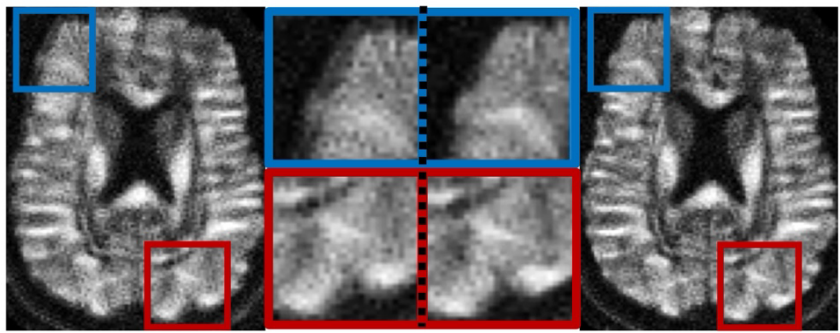

Figure 4. (a) The network's field estimation closely matches the ground truth spatiotemporal field (acquired via Skope). (b) Spatiotemporal eddy current and $dB_0$ field correction using field parameters estimated from the CNN; blur artifacts in certain areas are mitigated. (c–d) Field estimation and correction results from the unrolled ACE-Net. With the help of the data consistency term, the unrolled network outperformed the CNN in field parameter estimation.

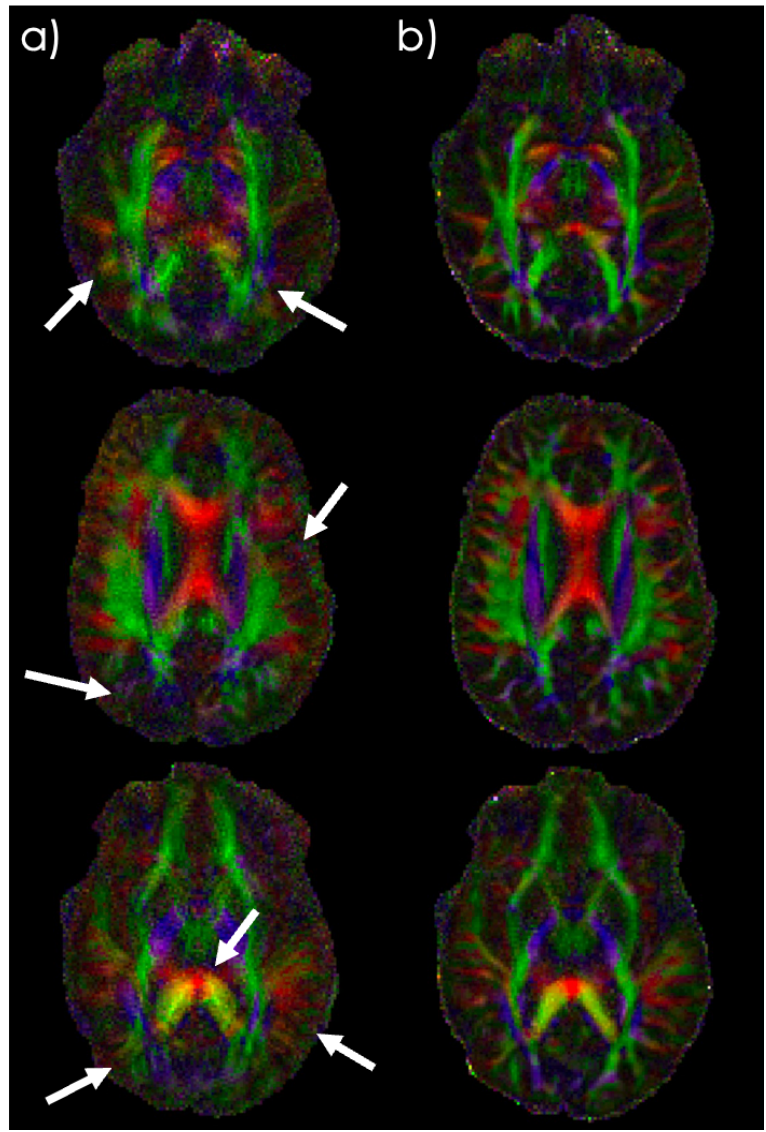

Figure 5. Diffusion Tensor Images (DTIs) reconstructed using 22-direction DWIs and a b = 0 image. (a) Reconstruction before $B_0$ and eddy current field estimation: the white matter fiber tracts are difficult to discern. (b) Reconstruction after $B_0$ and eddy current field estimation using ACE-Net: image quality is significantly improved, and the brain white matter fiber tracts, highlighted by arrows, become clearly visible after correction.

Acknowledgements
We would like to acknowledge the support from the National Institutes of Health (NIH) under grants R01MH116173, R01EB019437, U01EB025162, P41EB030006, R01EB033206, and U24NS129893. We also thank GE Healthcare for their support.